\documentstyle[epsfig]{l-schoolx}

\def\as{\alpha_S}
\def\asp{\frac{\alpha_S}{\pi}}
\def\shift  {\rule[-3mm]{0mm}{8mm}}
\newcommand{\beq}{\begin{equation}}
\newcommand{\eeq}{\end{equation}}
\newcommand{\bea}{\begin{eqnarray}}
\newcommand{\eea}{\end{eqnarray}}

\def\gev{\;{\rm GeV}}
\def\mev{\;{\rm MeV}}
\def\cO{{\cal O}}
\def\msb{\overline{{\rm MS}}}
     
\begin{document} 
 
\markboth{W.J. Stirling}{$\as$: FROM DIS TO LEP}
  
\setcounter{part}{0} 
 
\begin{flushright}
DTP/97/80 \\
June 1997 \\
\end{flushright}

\title{\boldmath{${\as}$}: from DIS to LEP\footnote{Based on a talk
presented at the `New Non-Perturbative Methods and Quantisation
on the Light Cone' conference, Les Houches, France, March 1997}} 
 
\author{W.J. Stirling} 

\institute{Departments of Mathematical Sciences and Physics,\\ 
                    University of Durham,   Durham DH1~3LE,  England
          } 
\maketitle   
  
\section{INTRODUCTION}
\label{sec:intro}
The strong coupling $\as$ is a fundamental parameter of the Standard
Model. In comparison to parameters like $\alpha_{\rm em}$, $M_Z$
and $\sin^2\theta_W$ it is relatively poorly known. However the 
precision of $\as$ measurements
has improved dramatically in recent years. More than twenty
different types of process, from lattice QCD studies to the 
highest energy colliders, can be used to measure $\as$ accurately.
The most precise determinations now quote uncertainties in $\as(M_Z^2)$
of less than $5\%$. There is also a remarkable consistency between
the various measurements.

A comprehensive review of $\as$ measurements, including detailed descriptions
of the underlying physics for the most important processes, can be found
in Ref.~\cite{BOOK}. One year later, several of the measurements
quoted in Ref.~\cite{BOOK} have been updated, resulting in a slight shift
in the overall `world average' value. The purpose of the present review
is to update the discussion on $\as$ measurements  given in Ref.~\cite{BOOK},
focusing on the new values reported in the last year. For more theoretical
details, descriptions of other measurements and a full set of references,
the reader is referred to the original review in Ref.~\cite{BOOK}. 

\begin{figure}[htb]
\begin{center}
\mbox{\epsfig{figure=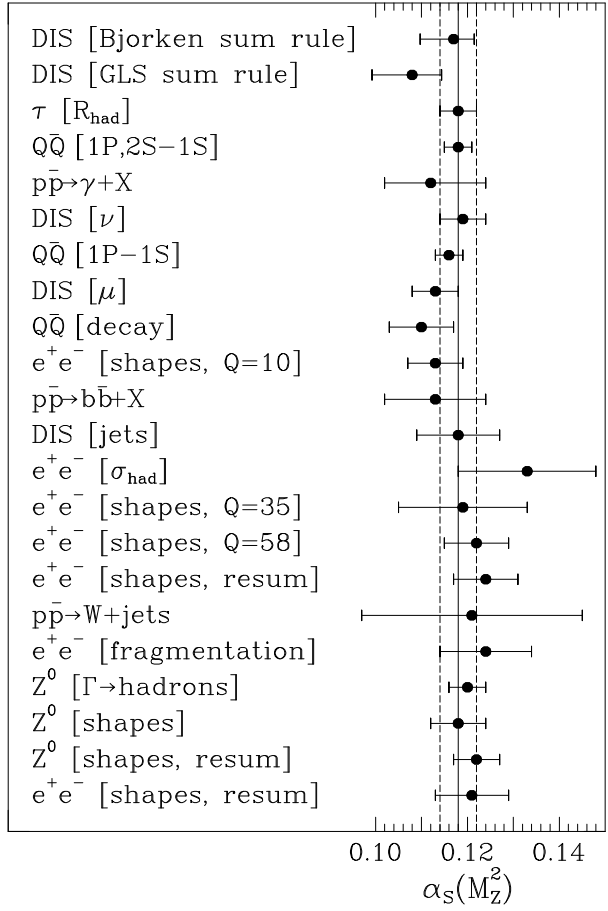,height=14cm}}
\caption{Measurements of $\as(M_Z^2)$, in the $\protect\msb$
renormalisation scheme,  updated from 
Ref.~\protect\cite{BOOK}.}  
\end{center}
\label{Fig1}  
\end{figure}  
The current situation is summarised in Fig.~\ref{Fig1}, which updates
Table 12.1 of Ref.~\cite{BOOK}. Before discussing the new measurements
in detail, we
begin with some technical preliminaries. In perturbative QCD
the dependence of the strong coupling on the renormalisation scale
is determined by the $\beta$--function:
\bea
Q^2 { \partial \as(Q^2) \over \partial Q^2 } & = & 
\beta(\as(Q^2) ) ,  \nonumber \\
\beta(\as) & = & - b \as^2 \left( 1 + b'\as + b'' \as^2 + \ldots
\right)  ,
\label{eq:betadef}
\eea
where $b= (33-2n_f)/(12\pi)$ etc. The coefficients in the perturbative
expansion depend, in general, on the renormalisation scheme (RS),
although for massless quarks the first two coefficients, $b$ and $b'$,
are RS independent. In essentially all phenomenological applications
the $\msb$ RS is used;  see Ref.~\cite{BOOK} for further discussion and
explicit expressions for the known $\beta$--function coefficients.

At leading order, i.e. retaining only the coefficient $b$, 
Eq.~(\ref{eq:betadef}) can be solved for $\as$ to give
\beq
\as(Q^2) = {   \as(Q^2_0) \over  1 + \as(Q^2_0) \; b \ln(Q^2/Q_0^2) } 
\eeq
or
\beq
\as(Q^2) = {  1  \over  b \ln(Q^2/\Lambda^2) } .
\eeq
These two expressions are entirely equivalent -- they differ only
in the choice of boundary condition for the differential equation,
$\as(Q_0^2)$ in the first case and the dimensionful parameter
$\Lambda$ in the second.
In fact nowadays $\Lambda$ is disfavoured as the fundamental parameter
of QCD, since its definition is not unique beyond leading order (see below),
and its value depends on the number of `active' quark flavours.
Instead, it has become conventional to use the value
of $\as$ in the $\msb$ scheme at $Q^2 = M_Z^2$ as the fundamental
parameter. The advantage of using $M_Z$ as the reference scale is
that it is (a) very precisely measured  \cite{PDG},
(b)  safely in the perturbative regime, i.e. $\as(M_Z^2) \ll 1$, 
and (c)  far from quark thresholds, i.e. $m_b \ll M_Z \ll m_t$.

The parameter $\Lambda$ is, however, sometimes still used
as a book-keeping device. At next-to-leading order there are
two definitions of $\Lambda$ which are widely used in the 
literature:
\bea
\mbox{definition 1}:&\quad &  b\ln{Q^2\over \Lambda^2} =
\frac{1}{\as(Q^2)} + b' \ln\left( 
{ b' \as(Q^2) \over 1 + b' \as(Q^2) } \right),
 \\
\mbox{definition 2}:&\quad & 
\as(Q^2) = {1\over 
 b\ln(Q^2 / \Lambda^2) } \left[ 
1 - \frac{b'}{b} {\ln\ln(Q^2 / \Lambda^2) \over 
\ln(Q^2 / \Lambda^2 ) } \right] .
\eea
The first of these solves Eq.~(\ref{eq:betadef}) exactly
when $b''$ and higher coefficients are neglected, while the second
(the `PDG' definition \cite{PDG})
provides an explicit expression for $\as(Q^2)$ in terms of $Q^2/\Lambda^2$
and is a solution of Eq.~(\ref{eq:betadef}) up to terms of order
$1/\ln^3(Q^2/\Lambda^2)$. 
Note that these two $\Lambda$ parameters are {\em different} for the 
{\em same} value of $\as(M_Z^2)$, the difference being about one quarter
the size of the current measurement uncertainty:
\beq
\Lambda_1^{(5)} - \Lambda_2^{(5)} \simeq 15\mev \simeq \frac{1}{4}
\delta_{\rm exp} \Lambda^{(5)}.
\eeq

\begin{table}[htb]
\label{tab1}
\begin{center}
\caption{Summary of the most important processes for $\as$
determinations in $e^+e^-$
collisions and in deep inelastic lepton-hadron scattering.}
\medskip  
\begin{tabular}{|c|l|l|} \hline
\shift  & quantity & perturbation series  \\ \hline
\shift $e^+e^-$   & $R_{ee}$, $R_Z$, $R_\tau$  &  $R=R_0[1+\as/\pi + \ldots ]$ \\                                               
\shift    &    event shapes, $f_3$, $\ldots$ &
$ 1/\sigma d\sigma / d X = A \as + B\as^2 + \ldots $ \\
\shift  &  $D^h(z,Q^2) $  &  $\partial D^h/\partial\ln Q^2 = \as D^h \otimes P + \ldots$ \\
\hline
\shift  $\ell N$ DIS & $F_i(x,Q^2)$   &   $\partial F_i/\partial \ln Q^2
 = \as F_i \otimes P + \ldots $  \\
\shift & &  $\int dx F_i(x,Q^2) = A + B\as + \ldots $ \\
\shift &   $\sigma(2 + 1 \; \mbox{jet})$  & $\sigma = 
A \as + B \as^2 + \ldots $ \\
\hline
\end{tabular}
\end{center}
\end{table}
In this review we will be mainly concerned with measurements
from $e^+e^-$ colliders (in practice LEP and SLC) and from 
deep inelastic scattering. Both processes offer several essentially
independent measurements, summarised in Table~\ref{tab1}. Note that 
all of these use the $q \bar q g$ vertex to measure $\as$, with the
high $Q^2$ scale provided by an electroweak gauge boson, for example
a highly virtual $\gamma^*$ in DIS or an on-shell $Z^0$ boson at LEP1
and SLC. There are two  main theoretical issues which affect  these determinations. The first is  the effect of 
unknown higher-order (next-to-next-to-leading order (NNLO) 
 in most cases) perturbative
corrections, which leads to a non-negligible renormalisation
scheme dependence  uncertainty in the extracted $\as$ values. This is 
particularly true for the event shape measurements at $e^+e^-$
colliders. The exceptions here are the total $e^+e^-$ hadronic
cross section (equivalently, the $Z^0$ hadronic decay width)
and the DIS sum rules, which are known to NNLO. The second issue
concerns the residual impact of $\cO(1/Q^n)$ power corrections.
For some processes it can be shown that the leading corrections
are $\cO(1/Q)$ (for example $\cO(1/M_Z)$ for the  corrections
to event shapes at LEP1 and SLC) which can easily be  comparable
in magnitude to the  NLO perturbative contributions. In deep inelastic
scattering, the higher-twist power corrections are 
$\cO(1/Q^2(1-x))$ and must be included in  scaling violation
fits especially at large $x$. Such power corrections (and their
uncertainties) must be taken
into account in $\as$ determinations, either using phenomenological
parametrisations or theoretical models.

Before discussing the new high-energy collider measurements of
$\as$  it is important to mention also
determinations from lattice QCD, which have very small uncertainties.
One of the simplest ways to define $\as$ on the lattice is 
to use the average value of the $1\times1$ Wilson loop
(plaquette)  operator:
\beq
\ln {\rm W}_{1,1} = \frac{4\pi}{3} \alpha_P\left(\frac{3.4}{a} \right)
\; \left[ 1 - (1.19 + 0.07 n_f) \alpha_P \right] ,
\label{eq:lat1}
\eeq
where $a$ is the lattice spacing.
A variety of choices is available for determining  $a$, i.e.
measuring the scale at which $\alpha_P$ has the value measured
in (\ref{eq:lat1}). Quarkonium level splittings, for example
$\Upsilon(S-P)$ and $\Upsilon(1S-2S)$, are particularly suitable.
Subsequently the plaquette $\alpha_P$ can be converted to the
standard $\msb$ $\as$ for comparison with other determinations:
\beq
\alpha_S^{(\msb,nf)} (Q^2) = 
\alpha_P^{(n_f)}(e^{5/3}Q^2) \; \left[ 1 + \frac{2}{\pi}
\alpha_P^{(n_f)} + C_2(n_f) \left( \alpha_P^{(n_f)}\right)^2 + \ldots \right].
\eeq
At present the two-loop coefficient is known only
for $n_f = 0$ \cite{c2ref} -- the shift  in $\as$ between using
$C_2(n_f = 0)$ and $C_2 = 0$ can be used to define a `conversion' error.
Several new lattice $\as$ values have been obtained recently, see
for example Ref.~\cite{junko}, and are included in Fig.~\ref{Fig1}.
As an example of the high precision of these measurements, we quote
the value obtained by the NRQCD collaboration  \cite{NRQCD} 
using the $\Upsilon(S-P)$ splitting:
\beq
\alpha_S^{\msb} (M_Z^2) =
0.1175 \pm 0.0011 ({\rm stat.}+{\rm sys.})
       \pm 0.0013 (m_q^{\rm dyn.})
       \pm 0.0019  ({\rm conv.}),
\eeq
where the first error is due to the lattice statistics and systematics,
the second is from the extrapolation in the dynamical quark mass, and the third
is the conversion error mentioned above.

In the following sections we will discuss new $\as$ measurements from LEP/SLC
and from deep inelastic scattering.
 Section~\ref{sec:summary} presents a new value for the
$\as$ world average.

\section{$\as$ from LEP and SLD}
\label{sec:ee}
In principle the most reliable determination
of $\as$ at the LEP and SLD $e^+e^-$ colliders
comes from the $Z^0$ hadronic width. In particular
we have, for the ratio $R_Z$,
\beq
R_Z = 
\frac{\Gamma(Z^0\to \mbox{hads.})}{\Gamma(Z^0\to \ell^+\ell^-)}
= R_0 \left[ 1 + \asp + C_2 \left(\asp\right)^2
+ C_3 \left( \asp \right)^3 + \ldots \right] 
\eeq
with $R_0 = 3 \sum_q(v_q^2 + a_q^2) / (v_\ell^2 + a_\ell^2) $.
The perturbative coefficients are known up to third order,
see  Ref.~\cite{BOOK} for explicit expressions and references, 
and as a result the prediction is 
very stable with respect to variations in the renormalisation scale.
In practice, since $R_0$ depends on the weak mixing angle and  
other electroweak parameters, it is more appropriate to perform
a global fit to all relevant electroweak quantities, for example
the (LEP and SLD) $Z^0$ partial widths and decay asymmetries, $p \bar p$ collider 
measurements of $M_W$ and $m_t$, etc. Such analyses are performed
regularly by the LEP Electroweak Working Group, and the results of   
a recent (1996) fit \cite{LEPEWWG96} are summarised in Table~\ref{tab2}.
\begin{table}[htb]
\label{tab2}
\begin{center}
\caption{Values for the Standard Model parameters obtained
from a global fit to LEP, SLD, $p \bar p$ and  $\nu N$ data,
from Ref.~\protect\cite{LEPEWWG96}.}
\medskip  
\begin{tabular}{|c|c|} \hline
 parameter & fit value \\ \hline
 $m_t$ [GeV]  &   $172 \pm 6$   \\
 $M_H$ [GeV]  &   $149^{+148}_{-82}$   \\
 $\as(M_Z^2)$  &  $0.120  \pm 0.003$   \\ \hline
 $\sin^2\theta_{\rm eff}^{\rm lept}$  &   $0.23167 \pm 0.00023$   \\
 $1-M_W^2/M_Z^2$  &   $0.2235 \pm 0.0006$   \\
 $M_W$ [GeV]  &   $80.352 \pm 0.033$   \\
\hline
\end{tabular}
\end{center}
\end{table}
An additional theory error from unknown higher-order corrections
of $\delta\as(M_Z^2) = \pm 0.002$ has been estimated,
see Ref.~\cite{LEPEWWG96} and references therein. The resulting $\as$
value,  
\beq
\as(M_Z^2) = 0.120 \pm 0.003 ({\rm fit}) \pm 0.002 ({\rm theory}),
\eeq
is displayed in Fig.~\ref{Fig1}.

The other high-precision determination of $\as$ at LEP and SLC
comes from {\em event shapes}, quantities which measure the
relative contribution  of the $\cO(\as)$ $e^+e^-\to q \bar q g$
process to the total hadronic cross section, see Table~\ref{tab1}.
A typical example is the thrust distribution:
\beq
\frac{1}{\sigma}\; \frac{d\sigma}{d T} = 
\as A_1(T) + \as^2 A_2(T) + \ldots  + \cO\left( 
\frac{1}{E_{\rm cm}}\right) .
\eeq
Such quantities are known in perturbation theory to
$\cO(\as^2)$, and the theoretical predictions in the
$T\to 1$ region can be improved
by resumming the leading logarithmic $A_n \sim \ln^{(2n-1)}(1-T)/(1-T)$
contributions to all orders, as discussed in  Ref.~\cite{BOOK}.
Another important recent theoretical development has been an improved
understanding of the leading $\cO(1/E)$ power corrections \cite{PC}, which 
at LEP can be as numerically important as the next-to-leading
perturbative corrections. 

\begin{figure}[htb]
\begin{center}
\mbox{\epsfig{figure=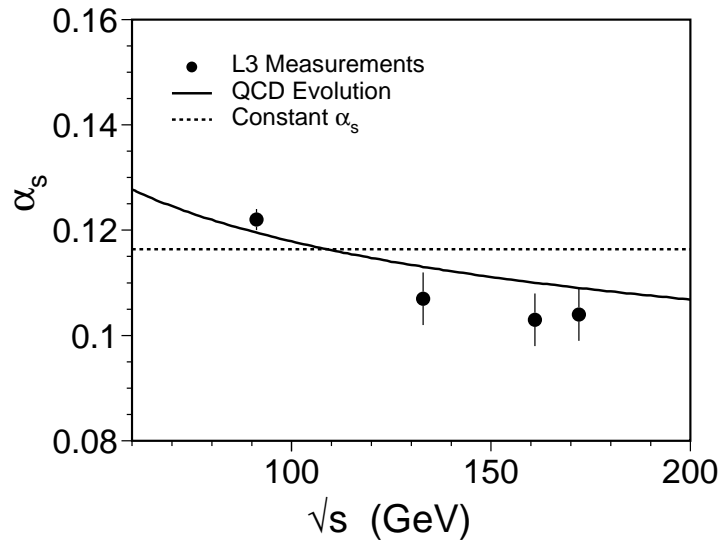,height=7cm}}
\caption{Measurements of $\as$ from event shapes at LEP1 and LEP2
from the L3 collaboration \protect\cite{L3}. The errors 
correspond to experimental uncertainties.}
\label{Fig3}  
\end{center}
\end{figure}

Event shapes have yielded $\as$ measurements over a wide range
of $e^+e^-$ collision energies, the most recent measurements being
at the LEP2 energies $\sqrt{s} = 161$ and $172\gev$. Although 
the statistical precision of these measurements cannot match
that obtained at the $Z^0$ pole, 
the results are consistent with
the $Q^2$ evolution of $\as$ predicted by Eq.~(\ref{eq:betadef}). For example,
Fig.~\ref{Fig3} shows the $\as$ values determined by the L3 collaboration
\cite{L3}
from event shape measurements at LEP1 and LEP2 energies. The solid line
is the evolution predicted by perturbative QCD.
Figure~\ref{Fig1} contains a new `LEP1.5'  average value for $\as$ obtained from event  
shapes at $\sqrt{s} = 133\gev$, taken from the 
1996 review by Schmelling \cite{SCHMELLING}: 
\beq
\as(Q^2=(133\gev)^2) = 0.114 \pm 0.007 \quad \Rightarrow \quad
\as(M_Z^2) =  0.121 \pm 0.008. 
\eeq
Another updated value in Fig.~\ref{Fig1} is that obtained from the scaling
violations of the fragmentation function measured in $e^+e^- \to hX$ over
a range of collision energies, the analogue of the scaling violations
of structure functions in DIS. The new value (an ALEPH/DELPHI average
taken from Ref.~\cite{SCHMELLING}) corresponds to 
\beq
\as(M_Z^2) =  0.124 \pm 0.010. 
\eeq

Finally, the CLEO collaboration have published \cite{CLEO} a new value for $\as$
obtained from the relative decay rate of the $\Upsilon(1S)$ into
a single hard photon:
\beq
\frac{\Gamma^{\gamma gg}}{\Gamma^{ggg}} = 
\frac{4}{5}\; 
\frac{\alpha}{\as(\mu^2)}\;
\left[ 1 -\left( 2.6 - 2.1 \ln(m_b^2/\mu^2)\right) \frac{\as(\mu^2)}{\pi} 
+ \ldots \right] .
\eeq
The new value, 
\bea
\as(M^2_{\Upsilon(1S)}) & = & 0.163 \pm 0.002({\rm stat.}) \pm 0.014 ({\rm sys.})
 \nonumber \\
  \Rightarrow \quad
\as(M_Z^2)& = & 0.110 \pm 0.001({\rm stat.}) \pm 0.007 ({\rm sys.}),
\eea
is included in Fig.~\ref{Fig1}.

\section{$\as$ from Deep Inelastic Scattering}
\label{sec:dis}
\begin{figure}[htb]
\begin{center}
\mbox{\epsfig{figure=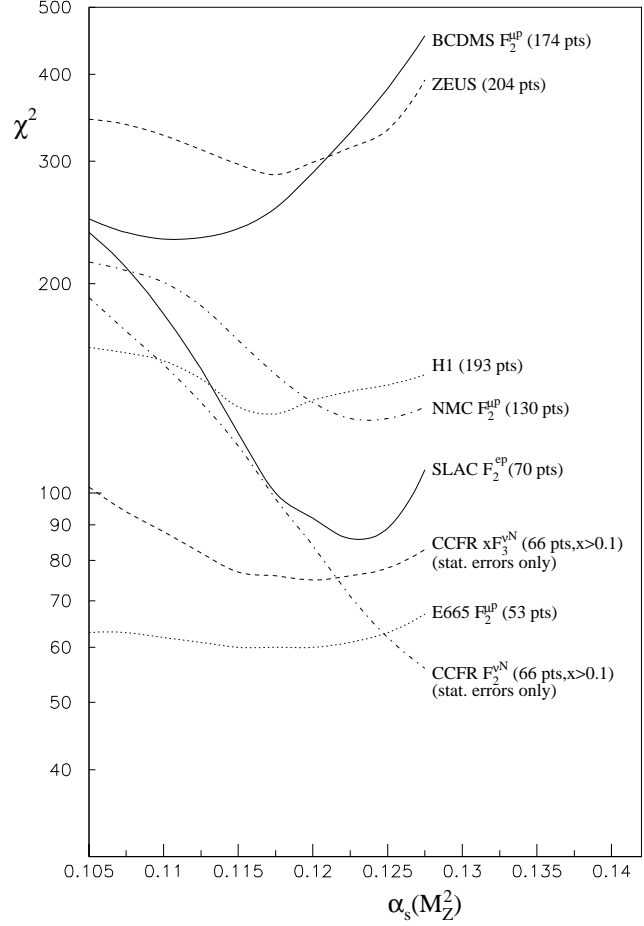,height=12cm}}
\caption{$\chi^2$ values for various DIS data sets obtained in a global fit to these 
and other hard scattering data \protect\cite{MRS}.}  
\label{Fig5}  
\end{center}
\end{figure}

The traditional method of measuring $\as$ in deep inelastic scattering
is from the strength of the structure function scaling violations 
predicted by the DGLAP equations:
\begin{eqnarray}
Q^2 {\partial q^{NS}\over \partial Q^2} 
                     &=& {\as(Q^2)\over 2 \pi} P^{qq} \otimes q^{NS} \nonumber\\
Q^2 {\partial q^S\over\partial Q^2} &=& {\as(Q^2)\over 2\pi} \left( P^{qq} 
\otimes q^S
+ 2 n_f P^{qg} \otimes g\right) \nonumber\\
Q^2 {\partial g\over\partial Q^2} &=& {\as(Q^2)\over 2\pi} \left( P^{gq} 
\otimes  q^S
+ P^{gg} \otimes g\right) ,
\label{glapeqns}
\end{eqnarray}
where $q^{NS}$ and $q^S$ are respectively non-singlet and singlet
combinations of quark distribution functions.
The fixed target and HERA structure function data, spanning a large range
in $x$ and $Q^2$, are all consistent with NLO DGLAP evolution, and yield
 $\as$ values which are in broad agreement. As an example, Fig.\ref{Fig5} \cite{MRS} shows
the $\chi^2$ values for various DIS data sets as a function of the $\as(M_Z^2)$
value  in the evolution equations. With one exception, all data sets 
exhibit a $\chi^2$ minimum in the $\as = 0.11 - 0.13$ range. In fact 
the `best fit' value for these data sets 
is $\as(M_Z^2)  = 0.118$, exactly the world average 
value (see Section~\ref{sec:summary} below). 

\begin{figure}[htb]
\begin{center}
\mbox{\epsfig{figure=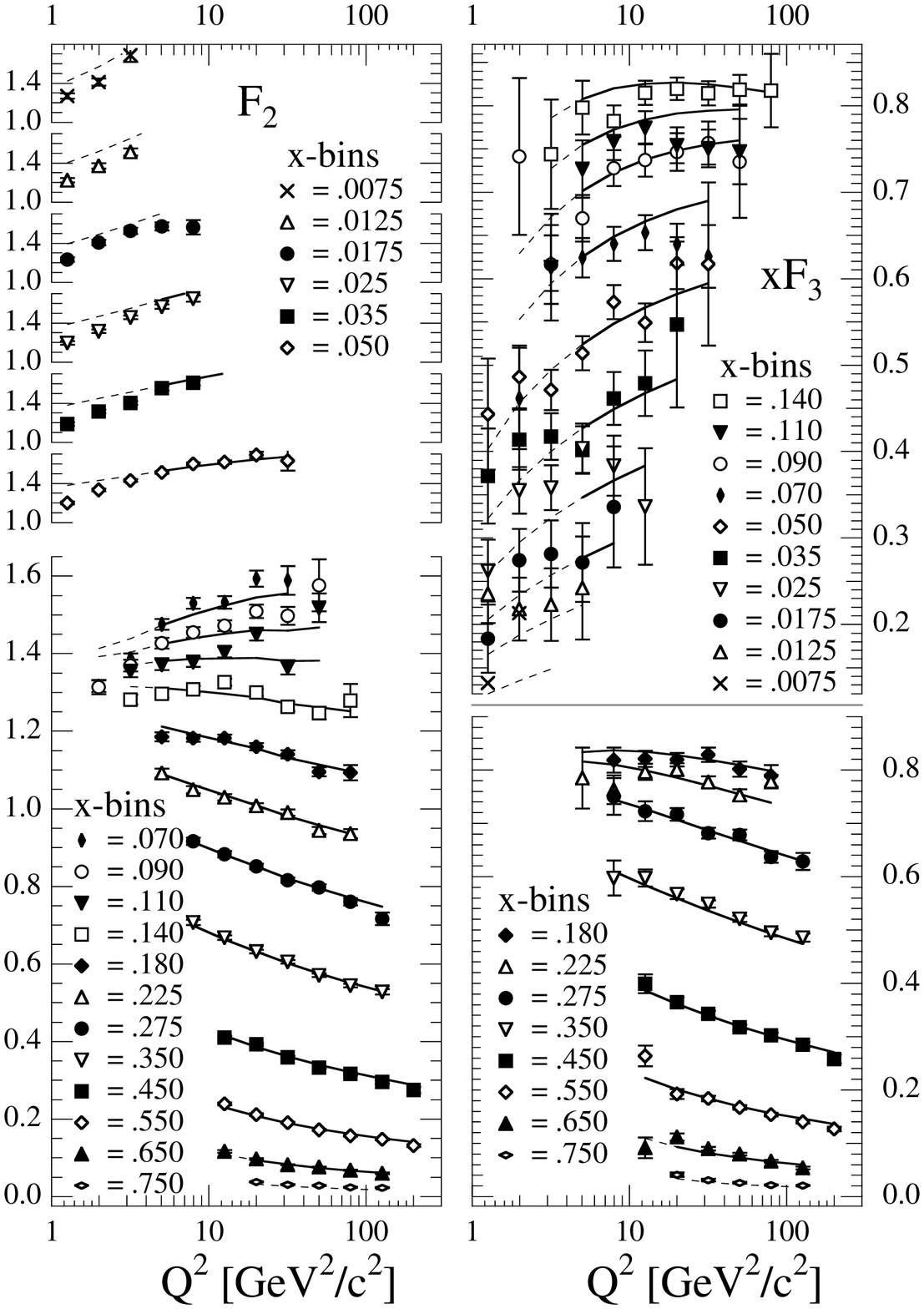,height=14cm}}
\caption{Measurements of the structure function $F_2^{\nu N}$ from the CCFR collaboration together with a NLO QCD fit, from Ref.~\protect\cite{CCFR}.}  
\end{center}
\label{Fig4}  
\end{figure}

It is difficult to extract a proper error on $\as$ from 
such global fit analyses. This requires a rigorous treatment of systematic
errors and inclusion of higher-twist contributions in the fit. Several groups
have performed such analyses. For example. 
the Milsztajn--Virchaux analysis of the SLAC/BCDMS ($eN,\mu N$) data
\cite{MV}  yields 
\beq
\as(M_Z^2) = 0.113 \pm 0.005,
\eeq
where the error includes statistical, systematic and scale dependence 
uncertainties. Recently the CCFR collaboration have  reported \cite{CCFR} a new value
of $\as$ from their $F_2^{\nu N}, xF_3^{\nu N}$ high-precision data 
(see Fig.~\ref{Fig4}):
\beq
\as(M_Z^2) = 0.119 \pm 0.002 ({\rm exp.})
\pm 0.001 ({\rm HT}) 
\pm 0.004 ({\rm scale}).
\label{eq:ccfr}
\eeq
The second error is from an estimate of the higher-twist contribution
using the model of Ref.~\cite{DW}, and the third is the scale dependence
uncertainty implemented as in Ref.~\cite{MV}. Note that the value in 
(\ref{eq:ccfr}) is somewhat larger than the earlier (1993) CCFR value
of $\as = 0.111 \pm 0.004$. The change is due to new energy calibrations
of the detector \cite{CCFR}. 

Deep inelastic scattering structure functions satisfy a variety of
{\em sum rules}, corresponding to the conservation of various
nucleon quantum numbers. In general the parton model values of
the sums have $\cO(\as)$ corrections,  which can be used to
extract $\as$ from measurements of  structure function integrals
at  fixed $Q^2$. Two sums rules which have been used
to obtain precision measurements are the Gross--Llewellyn Smith
and Bjorken sum rules (see Ref.~\cite{BOOK} for more
discussion and references):
\bea
\mbox{GLS}: \quad \int_0^1 dx(F_3^{\nu p} + F_3^{\bar\nu p})
 & = & 6 \left[ 1 + \asp + \ldots \right] + \Delta_{\rm HT}, \\
\mbox{BjS}: \quad \int_0^1 dx(g_1^{p} - g_1^{n})
 & = & \frac{1}{6}\frac{g_A}{g_V}  \left[ 1 - \asp + \ldots \right]+ \Delta_{\rm HT}, 
\label{eq:bjs}
\eea
where $\Delta_{\rm HT}$ represents $\cO(1/Q^2)$ higher-twist contributions.
A new analysis \cite{EGKS} of polarised structure function measurements has 
produced 
an update of the $\as$ value from the Bjorken sum rule.
In Ref.~\cite{EGKS} Pad\'e Summation is used to reduce the 
theoretical error from the choice of renormalisation scheme 
in the calculation of  the perturbation series on the right-hand side of (\ref{eq:bjs}). The resulting theoretical error in $\as(M_Z^2)$ is estimated
at $\pm 0.002$:
\beq
\as(M_Z^2) = 0.117^{+0.004}_{-0.007}({\rm exp.})  \pm 0.002({\rm theory}).
\eeq
This new value is included in Fig.~\ref{Fig1}.

Finally, $\as$ can be obtained from jet fractions and event shapes
in DIS, see Table~\ref{tab1}. For example, NLO theoretical predictions are currently
being used at HERA to extract $\as$ from the relative rate of
`2$+$1' jet production  at high $Q^2$, the analogue of
$f_3$ in $e^+e^-$ annihilation. No new results have been
published since the review in Ref.~\cite{BOOK}.

\section{Summary}  
\label{sec:summary}
The average value\footnote{obtained by  $\chi^2$ minimisation, as
described in Ref.~\protect\cite{BOOK}} of the measurements presented in 
Fig.~\ref{Fig1} is 
\beq
\mbox{WORLD AVERAGE:} \qquad \as(M_Z^2)\;  =\;  0.118\;  \pm\; 0.004. 
\label{eq:world}
\eeq
Following Ref.~\cite{BOOK}, the error here is defined as `the uncertainty
equal to that of a typical measurement by a reliable method'.
In view of the recent improvements in the lattice,  $Z^0$ hadronic width, and DIS
($\nu N$) determinations, it seems appropriate to decrease
the uncertainty of $\pm 0.005$ in Ref.~\cite{BOOK} to 
$\pm 0.004$. The central value in (\ref{eq:world}) has increased
by $+0.002$ from that given in Ref.~\cite{BOOK}. This is due 
primarily to (a) increases of $+0.003$ and $+0.004$ in the 
central values of the two lattice
determinations, and (b) an increase in the CCFR $\nu N$ 
DIS scaling violation  central value of $+0.008$. In view of the remarkable
consistency of all the measurements, and in particular of those
with the smallest uncertainties, it seems unlikely that
future `world average' values of $\as$ will deviate significantly, if at all,
from the current value given in (\ref{eq:world}).
  
\ack{I am grateful to my co-authors Keith Ellis and Bryan
Webber for their help in preparing this review.
This work was supported in part by the EU Programme
``Human Capital and Mobility'', Network ``Physics at High Energy
Colliders'', contract CHRX-CT93-0537 (DG 12 COMA).}

\end{document}